\documentclass[preprint,notitlepage,amsmath,amssymb,aps, superscriptaddress]{revtex4-2}
\usepackage{amsmath}
\usepackage{xcolor}
\usepackage{graphicx}
\usepackage{placeins}
\begin{document}

\newcommand{\sulf}{SO\textsubscript{2}F} 
\newcommand{\hyd}{{H\textsubscript{2}}} 

\newcommand{\expectation}[1]{\langle #1 \rangle}
\newcommand{\bra}[1]{\langle #1 |}
\newcommand{\ket}[1]{| #1 \rangle}
\newcommand{\inner}[2]{\langle #1 | #2 \rangle}

\newcommand{\ann}[0]{\hat{a}}
\newcommand{\adag}[0]{\hat{a}^\dagger}

\title{Data-driven reactivity prediction of targeted covalent 
inhibitors using computed quantum features for drug discovery}

\date{\today}

\begin{abstract}
We present an approach to combine novel molecular features with experimental data within a data-driven pipeline.  The method is applied to the challenge of predicting the reactivity of a series of sulfonyl fluoride molecular fragments used for drug discovery of targeted covalent inhibitors.  We demonstrate utility in predicting reactivity using features extracted from a workflow which employs quantum embedding of the reactive warhead using density matrix embedding theory, followed by Hamiltonian simulation of the resulting fragment model from an initial reference state.  These predictions are found to improve when studying both larger active spaces and longer evolution times.  The calculated features form a \textit{quantum fingerprint} which allows molecules to be clustered with regard to warhead properties.  We identify that the quantum fingerprint is well suited to scalable calculation on future quantum computing hardware, and explore approaches to capture results on current quantum hardware using error mitigation and suppression techniques. We further discuss how this general framework may be applied to a wider range of challenges where the potential for future quantum utility exists.
\end{abstract}

\author{Tom W. A. Montgomery}
\affiliation{Hybrid Intelligence, Capgemini Engineering, Richmond House, Walkern Road, Stevenage, Hertfordshire SG1 3QP, U.K.}
\affiliation{Capgemini Quantum Lab, Capgemini, 147-151 Quai du Président Roosevelt,92130 - Ile-de-France, Issy-les-Moulineaux, France}
\author{Peter Pog\'{a}ny}
\affiliation{GSK Medicines Research Centre, Gunnels Wood Road, Stevenage, Hertfordshire SG1 2NY, U.K.}
\author{Alice Purdy}
\affiliation{Hybrid Intelligence, Capgemini Engineering, Richmond House, Walkern Road, Stevenage, Hertfordshire SG1 3QP, U.K.}
\author{Mike Harris}
\affiliation{Hybrid Intelligence, Capgemini Engineering, Richmond House, Walkern Road, Stevenage, Hertfordshire SG1 3QP, U.K.}
\author{Marek Kowalik}
\affiliation{Capgemini Quantum Lab, Capgemini, 147-151 Quai du Président Roosevelt,92130 - Ile-de-France, Issy-les-Moulineaux, France}\author{Alex Ferraro}
\affiliation{Hybrid Intelligence, Capgemini Engineering, Richmond House, Walkern Road, Stevenage, Hertfordshire SG1 3QP, U.K.}
\author{Hikmatyar Hasan}
\affiliation{Hybrid Intelligence, Capgemini Engineering, Richmond House, Walkern Road, Stevenage, Hertfordshire SG1 3QP, U.K.}
\author{Darren V. S. Green}
\affiliation{GSK Medicines Research Centre, Gunnels Wood Road, Stevenage, Hertfordshire SG1 2NY, U.K.}
\author{Sam Genway}
\affiliation{Hybrid Intelligence, Capgemini Engineering, Richmond House, Walkern Road, Stevenage, Hertfordshire SG1 3QP, U.K.}
\affiliation{Capgemini Quantum Lab, Capgemini, 147-151 Quai du Président Roosevelt,92130 - Ile-de-France, Issy-les-Moulineaux, France}

\maketitle
\newpage

\section{Introduction}

Simulation of physical quantum systems is widely established as future high-value use case for quantum computers~\cite{Cheng2020, Lloyd1996}. Insights gained into the physical properties of  molecules and materials have clear value for a wide range of applications where understanding the quantum nature of system is critical~\cite{Reiher2017, Cao2018}.  The areas where quantum computing will likely be of value are those for which classical techniques for quantum chemistry fail due to not being able capturing many-body correlations~\cite{Blunt2022, Lee2022, Santagati2023}. Given the constraints of quantum hardware, applications in the foreseeable future should focus on the smallest problem scales for which quantum computing offers a potential advantage over classical techniques.

Alongside the rapid growth of quantum computing research, there have been rapid advancements in the use of predictive and generative machine learning models across a broad range of molecular and material challenges~\cite{Dara2022, David2020, Sellwood2018, Paul2021}. Often these schemes uses classical descriptions such as molecular SMILES or graph representations~\cite{Segler2018, Pogany2018, Olivecrona2017, McCloskey2020, Jin2018}. However, for small datasets in particular, such models often exhibit poor ability to generalise out of the training set distribution~\cite{ji2022}.

This work seeks to exploit the best of both areas in making predictions about a series of molecules, through capturing representations based on many-body electronic structure in a data-driven pipeline.  Previous work has highlighted the success of data-augmented models in chemistry~\cite{McClean2021} and it has already been shown that data-driven models using electron density can show large transferability even with a limited training dataset \cite{Nagai2020}. Quantum machine learning algorithms that directly calculate and process the  quantum ground states of Ising and Heisenberg models \cite{Uvarov2019} have been used to predict quantum phase transitions.  Furthermore, a quantum evolution kernel protocol has been developed, which uses quantum dynamics to produce representations of classical graphs that are hard to produce using a classical algorithm~\cite{Henry2021}. 

In this work we look at developing and generalising these ideas and applying them to electronic Hamiltonians to explore whether it is possible to create perfomant models trained on small datasets.  A central challenge centres on the Hilbert space dimension required even for relatively small molecules.  Rather than look to approximate methods developed for classical computers, we use density matrix embedding theory~\cite{Wouters2016} to allow Hamiltonian simulation of a molecular fragment subsystem using quantum algorithms.  Evolving the fragment for different times and  measuring relevant observables allows feature vectors for each molecule to be created, which may be used for data-driven predictive modelling tasks.  We note that Hamiltonian simulation is a natural task for quantum computers, with recent results for the Ising model suggesting state-of-the-art classical methods have been surpassed~\cite{Kim2023}.  As quantum hardware scales, we anticipate it will eventually become possible to apply the approach to increasingly large embedded fragments, surpassing scales which can be tackled with classical hardware, thus admitting the potential for quantum utility.

Targeted covalent drugs~\cite{Boike2022, Sutanto2020} have seen significant growth in research interest recently~\cite{Singh2022}. In contrast to traditional small-molecule inhibitors, modification of protein targets occurs through two steps: first, the drug molecule and target protein bind in a reversible reaction; second, a reactive group (known as a “warhead”) on the drug molecule reacts with an amino acid containing nucleophilic group on the target protein to form a covalent bond. The covalent binding can increase the potency, leading to correspondingly smaller doses, while additionally offering high selectivity. While historical discoveries date back all the way to aspirin, the recent growth in interest targeted covalent drugs in pharmaceutical discovery has been driven by advances in chemoproteomic assays enabling proteome-wide studies.

Drug discovery research has been progressively augmented and accelerated by \textit{in silico} methods, including both simulation and machine learning, with the latter playing an increasingly prominent role due to the growth in experimental data generated over time. Computational methods typically enable faster and larger explorations of drug-like molecules compared with in vitro experiments. A central challenge in the design of targeted covalent drugs is predicting the reactivity of the reactive warhead, which is critical to balancing properties such as potency and selectivity~\cite{McAulay2022}. Effective computational techniques to this challenge will have a high impact.

A particular chemical series which has been studied extensively with regard to  reactivity and other properties~\cite{Gilbert2023} consists of sulphonyl fluoride (\sulf{}) warheads.  It has been shown that reactivity predictions were possible using density functional theory (DFT) calculations of the lowest occupied molecular orbital (LUMO) energy. In contrast, other warheads, such as acrylamides, have required extensive DFT calculations for transition states in order to rank molecules’ reactivity in a way which is approximately consistent with experimental data~\cite{Lonsdale2017}. Classical machine learning approaches using classical molecular features as inputs have shown a degree of success in their ability to make approximate reactivity predictions at significantly lower computational cost~\cite{Palazzesi2020}.

The quantum mechanical nature of warhead reactivity mechanisms prompts exploration on whether quantum computational approaches could prove fruitful for reactivity prediction. The use of future quantum computers is motivated by a desire to use a fully quantum mechanical description of the molecules, including the potential to capture strongly correlated behaviour. At the same time, this work seeks to create a general end-to-end approach which could leverage experimental data alongside features generated from a quantum computer as part of a machine learning pipeline.

This work explores the sulfonyl fluoride chemical series detailed in Ref.~\cite{Gilbert2023} for which experimental data was already available to us.  It was chosen as a first paradigmatic example to explore with combined quantum computational and data-driven approaches for predicting warhead reactivity.  Although other challenges in drug discovery with multireference character will ultimately offer the potential for greater benefits from quantum computers, the sulfonyl fluoride chemistry was selected for multiple reasons: firstly, the availability of experimental and existing DFT calculations; secondly, the simpler reaction mechanism for sulfonyl fluoride warheads allows the new approach to be explored end-to-end with smaller active spaces, while still capturing real-world relevance; and thirdly, the opportunity to gain greater insights for this sulfonyl fluoride warhead.

This manuscript is organised as follows: in Section~\ref{sec:pred_models}, we give describe the end-to-end approached in this work.  Section~\ref{sec:pipeline_optimisations} discusses optimisations to the end-to-end pipeline.  Quantum algorithms and results captured on quantum hardware are presented in Section~\ref{sec:quantum_algos} before giving our conclusions.

\section{Predictive modelling pipeline with computed quantum features}
\label{sec:pred_models}

\begin{figure}[!hbt]
\includegraphics[scale=0.45]{"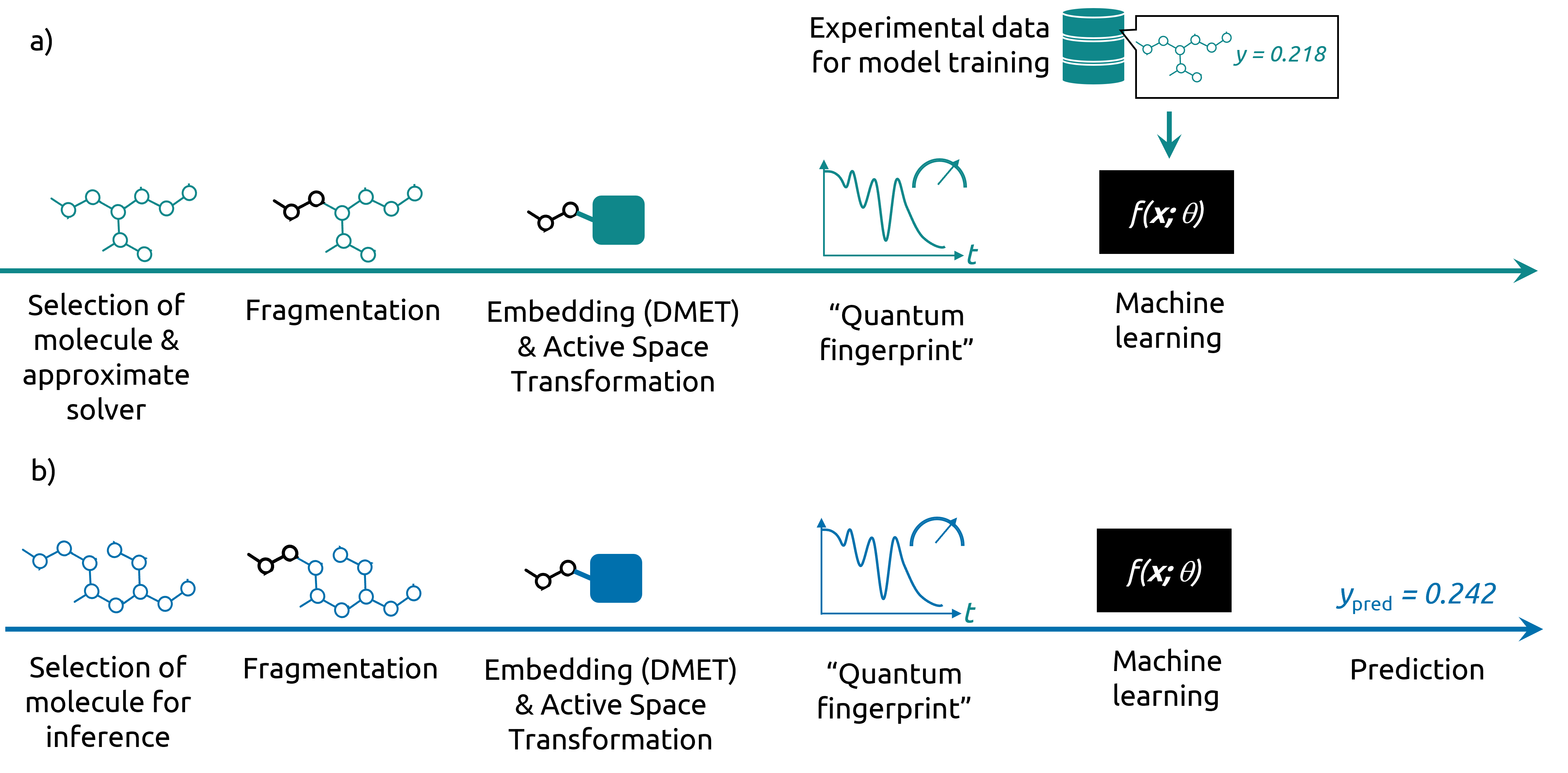"}
\caption{a) A high-level schematic indicating the end-to-end workflow for training predictive models which leverage computed quantum features. A fragmentation scheme is identified to find a relevant fragment which will be captured using a Hamiltonian found using DMET. Features are extracted from the fragment model through preparing the system in an initial state, applying a unitary transformation and measuring observables.  The resulting observables, which form a ``quantum fingerprint'', are combined with experimental data for the measurement to be predicted, and a machine learning model is trained.  b) The equivalent pipeline for inference is similar, with the same process for embedding and extraction of observables for the quantum fingerprint as for model training, but with the machine learning model used for inference.}
\label{fig:pred_models:end-to-end}
\end{figure}

Our high-level approach involves finding a relevant subsystem within a molecular system and obtaining an effective Hamiltonian describing a fragment active space and its entanglement with the rest of the system.  A feature vector is created from measuring observables $\expectation{O}$ which arise from preparing the subsystem in a particular initial state and transforming using a unitary $U_t(H)$ which depends on the many-body DMET Hamiltonian $H$.  The feature vector, which we refer to as a ``quantum fingerprint'', is used to train a machine learning model to predict a measurement of interest using a training dataset where measurements have been captured experimentally.  To make predictions for new molecules, the feature extraction steps are applied to the new molecules and the machine learning model is used for inference.  The approach is summarised in Fig.~\ref{fig:pred_models:end-to-end}.

The motivation for our approach is three-fold: firstly, the use of embeddings enables us to tackle challenges on different scales, with a view to obtaining predictive features on future quantum computers which are inaccessible on classical hardware.  Secondly, use of a machine learning pipeline affords {\it in silico} experiments to be performed without a need to simulate a larger interacting system explicitly.  This can be seen as a ligand-based drug discovery approach where  quantum features more closely linked to the predictive modelling task are leveraged -- thus offering the potential for predictive models which offer better performance and better generalisation to novel chemistry for a smaller set of training molecules.  Thirdly, there is great scope to optimise the pipeline for quantum hardware, since the transformations $U_t(H)$ may not need to be physically motivated for them to yield informative features for machine learning.  There is similar flexibility in the selection of initial states and observables which are measured, which can allow for chemical insight to drive the selection of features captured for the machine learning task.  Furthermore, we expect greater robustness to deterministic errors on quantum hardware because a machine learning model may be able to learn from systematically incorrect input data.

In the following subsections, we describe the core parts of the pipeline shown in Fig.~\ref{fig:pred_models:end-to-end} in detail, starting with the approach to embedding models which capture a relevant active space for molecules of interest.  We then describe the approach to extracting features from the fragment active space via initial state preparation, transformations $U_t(H)$ and measurements observables.

\FloatBarrier
\subsection{Embedding molecular fragments}
In this subsection, we explore quantum embedding models which allow the dynamics of molecular subsystems to be investigated.  First, the general approach to active space transformations used in this work is discussed, before moving to the challenge of creating embedding models of spatially local fragments using DMET.

The first step is to employ a self-consistent field (SCF) approach such as Hartree-Fock (HF) theory or Kohn-Sham density functional theory (KS-DFT). In this work we focus on HF approaches. We first express the ground-state wavefunction as a single Slater determinant of molecular orbitals $\ket{\phi_r}$.
The total electronic energy is then minimized, subject to an orbital orthogonality constraint; this is equivalent to the description of the electrons as independent particles that only interact via each others’ mean field.
Molecular orbitals are expressed as linear combinations of atomic orbitals which leads to the Roothaan equations, in matrix form: $\mathbf{F}\mathbf{C} = \mathbf{S}\mathbf{C}\mathbf{E}$.
Here $F$ is the Fock matrix, $\mathbf{C}$ is the matrix whose columns are the coefficients of the molecular orbitals in the atomic orbital basis, $\mathbf{S}$ is the matrix of atomic orbital overlap integrals and $\mathbf{E}$ is the diagonal matrices of molecular orbital energies.
The Fock matrix is made up of the sum of four terms $\mathbf{F} = \mathbf{T} + \mathbf{V} + \mathbf{J} + \mathbf{K}$ where $\mathbf{T}$ is the kinetic energy matrix, $\mathbf{V}$ is the external potential, $\mathbf{J}$ is the Coulomb matrix, and $\mathbf{K}$ is the exchange matrix.
The groundstate wavefunction can be written as:
\begin{equation}
\ket{\Phi_{\text{HF}}} = \prod^{N}_{r=1}{\adag_{r}}\ket{\text{vac}}
\end{equation}
where the $N$ electrons occupy the lowest molecular orbitals, with $\adag_r$ ($\ann_r$) the creation (annihilation) operators for molecular orbital $\ket{\phi_r}$.

Before discussing the case of an active space for a spatially local fragment within the molecule, we consider an active space transformation at the Hartree-Fock SCF level following~\cite{Rossmannek2020}.  Here, the first step is to assume that all orbitals at the SCF level of approximation that are occupied below some energy are `frozen' with the one-body density matrix remaining diagonal with double occupation in the SCF molecular orbital basis. The second step is to remove corresponding virtual orbitals above some energy threshold.  A procedure of tracing out inactive orbitals shifts the single-electron energies in the remaining active orbitals which results in a Hamiltonian of the form:
\begin{equation}
H^A=\sum^{\text{A}}_{rs}  h_{rs}^{\text{eff}} \adag_r \ann_{s} + \sum_{pqrs}^{\text{A}} (pq|rs) \adag_{p}\adag_{r} \ann_{s}\ann_{q}
\label{eq:H_active_space}
\end{equation}
where the sums run over the active space and $(pq|rs)$ are two-electron integrals and one-electron integrals are modified through interaction with the inactive electrons with $h_{rs}^{\text{eff}} = h_{rs} + \sum_i^\text{I} \left[2 (ii|rs) - (ir|si)\right]$ where the sum is over the inactive space and we have used the fact that the one-body density operator is diagonal in the molecular basis with entries of 2 or 0 depending on whether the inactive orbitals are occupied or virtual.  In this manuscript, we will refer to this as a HOMO-LUMO energy based active space transformation.

We now turn our attention to creating embeddings with spatial structure through the DMET approach, which may also be viewed as an active space transformation.  It is motivated by selecting a set of so-called `fragment' orbitals which should remain active in the embedded Hamiltonian.  In the DMET algorithm~\cite{Wouters2016}, a fragment is selected as a sub-space of molecular orbitals localised to a spatial region within the molecule -- for example, a particular functional group.  Fragment orbitals are selected from a complete orthonormal set of localised molecular orbitals.  Our strategy to create a localised set of molecules is to apply L\"owdin symmetric orthonormalisation of the atomic orbitals.  The complement to the fragment space is the `environment' space.  The DMET algorithm splits the environment into an active set of `bath' orbitals and an inactive set of occupied and virtual `inactive environment' orbitals.  

At the SCF level the inactive environment contains either approximately full or empty (virtual) orbitals and thus the number of electrons is approximately an integer $N_{\text{E}}$.  In the same manner used to derive the active space Hamiltonian $H^\text{A}$ (eq.~\ref{eq:H_active_space}), the occupied inactive orbitals are assumed to be frozen and enter the effective Hamiltonian of the active cluster as a shift in the energy of the cluster orbitals comprising the fragment and bath. Since the number of electrons in the inactive environment is approximately an integer at the SCF level, so is the number of electrons in the active cluster ($N_{\text{C}}$). This procedure leads to a fragment Hamiltonian of the form:
\begin{equation}
H=\sum^{\text{F}+\text{B}}_{rs}  h_{rs}^{\text{eff}} \adag_r \ann_{s} + \sum_{pqrs}^{\text{F}+\text{B}} (pq|rs) \adag_{p}\adag_{r} \ann_{s}\ann_{q} - \mu\sum_{r}^{\text{F}} \adag_{r}\ann_{r}
\label{eq:H_DMET}
\end{equation}
where F is the fragment and B the effective bath, whose Hilbert space dimension equals that of the subsystem by the Schmidt decomposition. The effective one-electron integrals $h_{rs}^{\text{eff}} = h_{rs} + \sum_{ij} \left[ (rs|ij) - (rj|is) \right]D^{\text{E}}_{ij}$  include the modification to the single-electron integrals $h_{rs}$ due to the interaction with the environment via the one-body density operator for the environment $D^{\text{E}}_{mn}$.  $\mu$ is a chemical potential which is selected to ensure the number of electrons in the cluster and environment equals the total number of electrons in the molecule.  

Having created a Hamiltonian for the fragment with the DMET process of the form in eq.~\ref{eq:H_DMET}, it is possible to further reduce the size of the active space.  In this work, this is done by rotating the basis of the cluster from the fragment and bath orbitals to one which diagonalises the projection of the Fock matrix $\mathbf{F}$ on to the cluster space.  An active sub-space within the DMET cluster is found as described above for eq.~\ref{eq:H_active_space}.  It should be noted that this is one of various possible approaches to identify an active subspace for the fragment.  For example, one could start with a smaller fragment and apply the DMET approach and follow this by appending other orbitals to the bath, such as some bond localized orbitals.  This latter approach, which may provide a more consistent approach to reducing the active space size, is not considered in this work.

\subsection{Quantum fingerprint calculations}

\begin{figure}[!hbt]
\includegraphics[scale=0.45]{"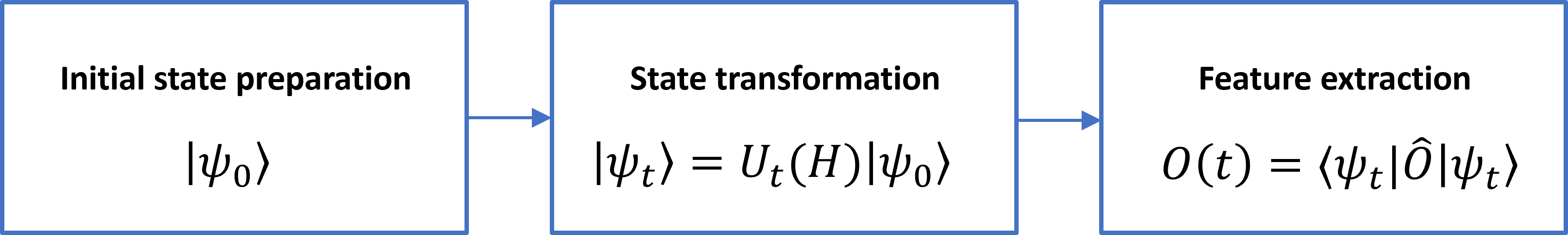"}
\caption{a) A high-level schematic indicating the workflow for creating a quantum fingerprint for downstream predictive modelling, including initial state preparation followed by application of a parametrised unitary operator and estimation of observables for different parameter values $t$.  }
\label{fig:pred_models:quantum_fingerprint}
\end{figure}

The approach to creating a feature vector, or \textit{quantum fingerprint}, involves state preparation, state transformation and extraction of features from quantum measurements of observables.  This naturally aligns with calculations which can be performed using quantum circuits on gate-based quantum devices, as illustrated in Fig.~\ref{fig:pred_models:quantum_fingerprint}.

For most of this work, we focus on Hamiltonian simulation and choose the state transformation
\begin{equation}
\ket{\psi_t} = U_t(H)\ket{\psi_0} = e^{-iHt}\ket{\psi_0}
\label{eq:evolution}
\end{equation}
where the parameter $t$ is explicitly the evolution time under the effective fragment Hamiltonian $H$.  Hamiltonian simulation is a natural task for quantum computers with polynomical scaling in $t$ and system size possible with straightforward approaches such as the Trotter-Suzuki method, with even better asymptotic scaling for post-Trotter methods involving qubitisation.  We will consider the approach to definiting $U_t(H)$ to be more general and $t$ may be considered to parametrise another unitary which encodes $H$.  This will be discussed in more detail in Section~\ref{sec:quantum_algos}, where different approximations to $e^{-iHt}$ are considered and we consider that in the more general context of the data-driven pipeline in Fig.~\ref{fig:pred_models:end-to-end}, $U_t(H)$ maybe \textit{defined} by a scheme which approximates unitary evolution under $H$.

\subsection{Pipeline results for sulfonyl fluoride warhead reactivity}
\label{sec:motivational_results}
We now present results which were obtained using noiseless quantum simulators, with quantum algorithms and quantum hardware results presented in Section~\ref{sec:quantum_algos}.  We  turn our attention to the central challenge of this manuscript -- the prediction of reactivity for chemical series with \sulf{} warheads.  Our work centres on a series of 100 molecules~\cite{Gilbert2023}, of which 8  have experimental reactivity measurements.  The remaining molecules have estimated relative reactivities from DFT calculations performed using B3LYP-D3 functional with basis set 6-31+G**.  The DFT calculations will be used as a proxy for experimental measurements.  While this precludes the possibility of improving predictions, or indeed future quantum utility on this dataset without capturing more experimental data, it will allow demonstration of the approach which assumes relatively little prior knowledge about the reaction mechanism.  

\begin{figure}[!hbt]
\includegraphics[scale=0.45]{"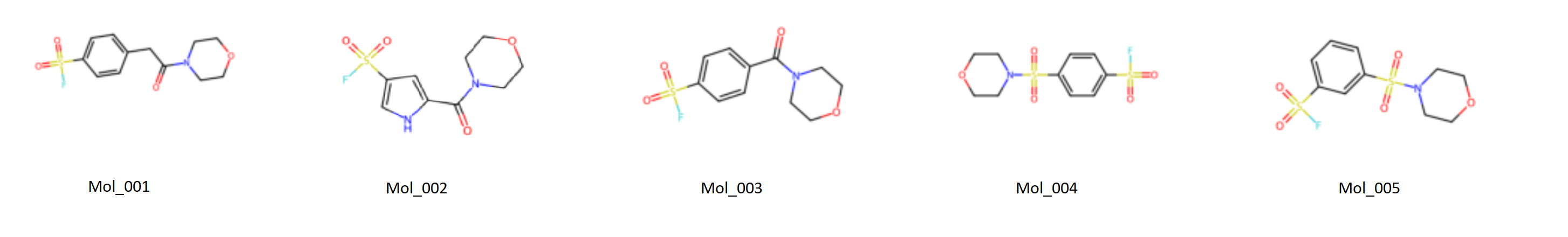"}
\caption{Examples of sulfonyl fluoride molecules in the data set.  The \sulf{} group, common across all molecules, is selected as a fragment subsystem for DMET.}
\label{fig:sulf:mols}
\end{figure}

Owing to the size of the molecules in the \sulf{} chemical series, we apply the DMET procedure to capture a local fragment and its interaction with the rest of the molecule.  For the example studied in this work, this is a natural choice of fragmentation strategy, which involves identifying the \sulf{} group itself as the fragment.  This is motivated by this part of the molecule containing the reaction centre for covalent binding to target residues.  In addition, that this group exists across all molecules in the chemical series (see Fig.~\ref{fig:sulf:mols} for example molecules) allows the capture of quantum fingerprint feature vectors on equivalent Hilbert subspaces on each molecule, allowing a natural comparison between the features.  

To calculate a feature vector for each sulfonyl fluoride warhead, we evolve the fragment under $H$ from an excited state
\begin{equation}
\ket{\Phi^\text{ex}} = \adag_{l, \uparrow} \adag_{l, \downarrow} \ann_{h, \uparrow} \ann_{h, \downarrow} \ket{\Phi_\text{HF}}\, .
\label{eq:initial_state}
\end{equation}  
Here, the subscripts $l$ and $h$ indicate lowest unoccupied and highest occupied fragment molecular orbitals, respectively, arrows indicate spin, and $\ket{\Phi_\text{HF}}$ is the highest Hartree-Fock fragment ground state.  This choice is motivated by previous work~\cite{Gilbert2023} demonstrating the utility of the molecular LUMO energy for predicting reactivity.  The observables captured at different times $t$ are elements of the one-body density matrix of the fragment $\rho_{rs}^\text{}(t) = \bra{\psi(t)}\adag_s \ann_r\ket{\psi(t)}$, where  $r$ and $s$ represent localised orbitals in the warhead fragment only.  The number of one-body density matrix elements grows quadratically with the dimension of the fragment.  To focus on a single temporal scalar, we propose following following temporal observable 
\begin{equation}
F(t) = \sum_{r,s}h_{rs}^\text{eff}\rho_{rs}(t)\, .
\label{eq:F_t}
\end{equation}
This means that dynamics in the density matrix are more heavily weighted if there is a stronger coupling between  two orbitals, either through the effective one electron terms or the direct coulomb interaction.  Throughout this manuscript, our results will be presented in energies relative to the Hartree energy $E_H$ and Hartree time $t_H (\simeq 0.024fs)$.

\begin{figure}[!hbt]
\includegraphics[scale=0.7]{"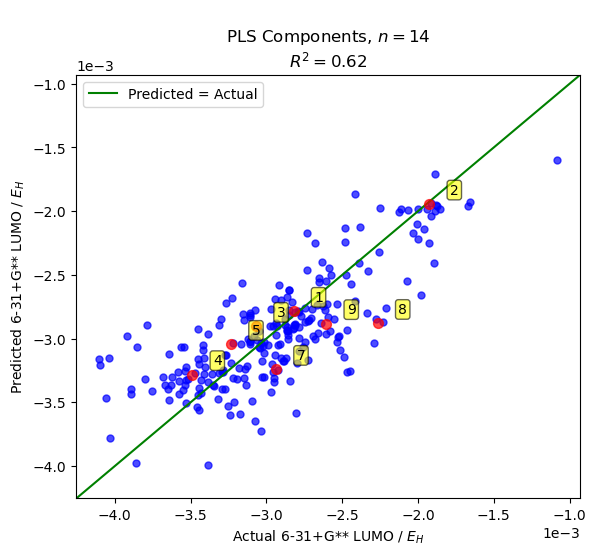"}
\caption{Cross-validation results for a PLS model trained using $F(t)$ and the LUMO energy from DFT calculations as a proxy for reactivity. Each point shows a predicted value from a validation set not used in model training as part of a 5-fold cross-validation scheme. Points highlighted in red are molecules with experimental reactivity data.  Results were generated for a (4e, 4o) active space for the initial state $\ket{\Phi^\text{ex}}$.}
\label{fig:sulf:pred_vs_actual}
\end{figure}

A partial least squares model (PLS) was trained using $F(t)$ found from statevector simulation for $t/t_H \in [0,14]$ with points sampled every 0.5$t_H$.  Data for the target variable, warhead reactivity, is taken in proxy as the LUMO energy found from DFT calculations.  Using a 5-fold cross-validation scheme, we select a PLS model with 14 components with a cross-validation explained variance of $R^2 = 0.61$.  Consistent performance on the subset of molecules with experimental reactivity is found, which span across the range of reactivities in the wider data set, as shown in Fig.~\ref{fig:sulf:pred_vs_actual}.

\begin{figure}[!hbt]
\includegraphics[scale=0.6]{"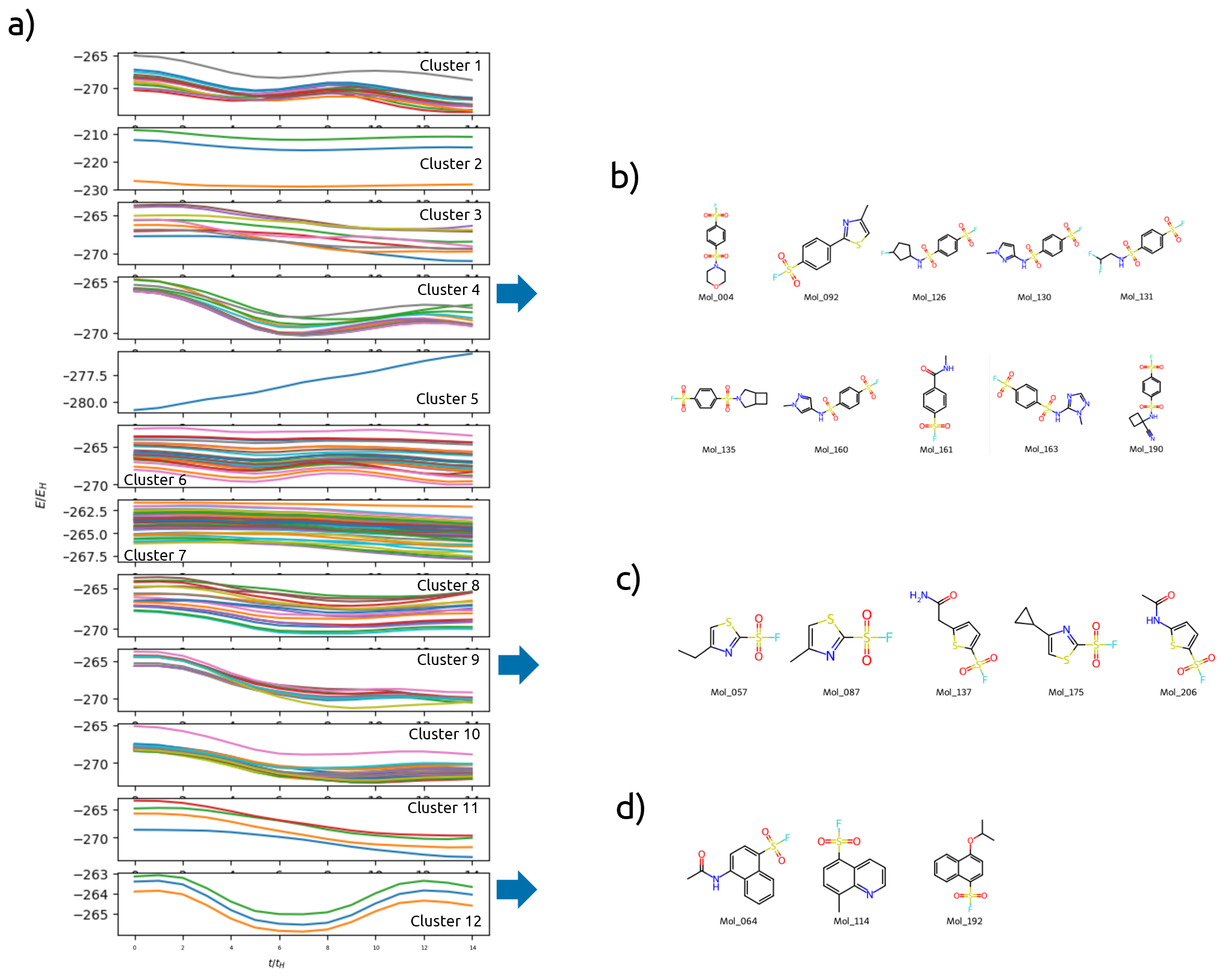"}
\caption{a) Clustering results for the features defining the quantum fingerprint.  Time series clustering was performed using the \textit{tsfresh} package to extract time series features. PCA was then employed to reduce the dimension of the feature set, with clustering performed using k-means.  The quantum fingerprints for molecules within each cluster are shown on separate, labelled plots. Results were captured for the (6e, 6o) active space with initial state $\ket{\Phi^\text{ex}}$. b) Example molecules from cluster 4. c) Example molecules from cluster 9. d) Example molecules from cluster 12.}
\label{fig:sulf:clustering}
\end{figure}

We close this section by looking at the nature of the quantum fingerprints defined in eq.~\ref{eq:F_t} for the sulfonyl fluoride molecular series, for the initial state defined in eq.~\ref{eq:initial_state}. Specifically, we look at whether the quantum fingerprints extracted from warhead dynamics allow the molecules to be clustered.  Fig.~\ref{fig:sulf:clustering} shows an effective clustering of the molecules based on the temporal features (see caption for details).  Example molecules from different clusters are shown, demonstrating that this approach is able to identify similar structures solely from electronic dynamics of the warhead from a prepared reference initial state.  For example, cluster 4 is mainly comprised of 1,4-substituted benzene rings with sulfonamide groups at the 4 position.  Cluster 9 includes thiazoles and thiophenes, and cluster 12 comprises naphthalenes.  Interestingly, the method is able to identify molecules with different structures but similar properties -- for example, Mol\_161 in cluster 4 with an amide in place of a sulfonamide group.

\FloatBarrier
\section{Pipeline optimisations}
\label{sec:pipeline_optimisations}
The positive results in the previous section lead to a number of questions that we look to address in this work. The ability to predict \sulf{} reactivity with modest errors demonstrated in Fig.~\ref{fig:sulf:pred_vs_actual} was found straightforwardly without exploration of different functional forms of $F(t)$, the underlying unitary $U_t(H)$, or even any detailed consideration of the evolution time (beyond it being a multiple of the inverse Hartree energy).  This prompts a question around how performance can be optimised and quantum computational resource minimised through effective choices of: initial state; measurement operators, evolution time and active space size and; potentially,  the choice of unitary $U_t(H)$ which encodes the Hamiltonian.  In the next subsection, the impact of varying the evolution time and active space size is studied; other alterations to the scheme are explored in Sections~\ref{subsec:initial_states} and~\ref{subsec:measurements}.

\FloatBarrier
\subsection{Evolution time and active space size}

\begin{figure}[!hbt]
\includegraphics[scale=0.6]{"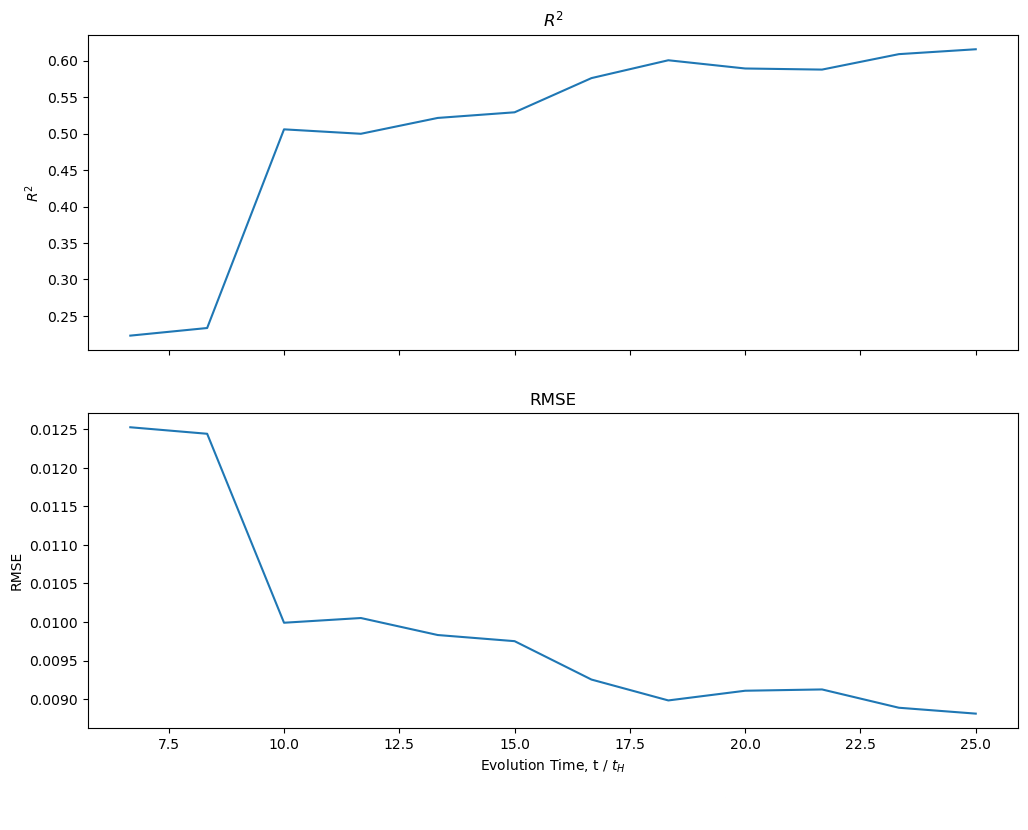"}
\caption{Model performance from the same 5-fold cross-validation approach in Fig.~\ref{fig:sulf:pred_vs_actual} for different time evolution lengths, indicating a step improvement in model performance when time scales reach $t/t_H\simeq 10$. Shown is the root mean square error (top), explained variance $R^2$ (bottom).  In each case, the initial state was $\ket{\Phi^\text{ex}}$ and the active space was (4e, 4o). }
\label{fig:sulf:evolution_times}
\end{figure}

Since the time evolution of the quantum states is the main component of the workflow to be performed on quantum computers we want to minimise the total required evolution time. In seeking to explore the question as to what length of time evolution is required to capture a quantum fingerprint which enables reactivity prediction to be possible, we repeated the approach when generated the results in Fig.~\ref{fig:sulf:pred_vs_actual} with a range of different times.  The results are shown in Fig~\ref{fig:sulf:evolution_times}, which shows the model performance improving approximately monotonically with increasing evolution time.  A dramatic improvement in predictive performance occurs when the time scale reaches $10t_H$ which corresponds to around 0.24 fs.  This suggests that the evolution needs to be sufficiently long to capture dynamics due to energy scales on the order of one tenth of a Hartree energy which is consistent with typical S-F bond energies in sulfonyl fluoride compounds.

\FloatBarrier 

In general, we expect the duration of evolution required to be a function of the initial state prepared and the choice of measurement operators.  These will be addressed in the next two subsections.  First, we turn our attention to the active space size used for the computation.

Let us consider another aspect relevant for implementation on quantum computers: the size of the active space. As mentioned in Section~\ref{sec:pred_models}, we perform an HOMO-LUMO energy based active space transformation to the cluster found from the DMET algorithm, which allows different active space sizes to be considered for the same fragmentation. The results in Section~\ref{sec:motivational_results} were based on the (4e, 4o) active space.  While there are quantum computers with over 100 qubits available today, the depth of circuits to approximate Hamiltonian evolution have a depth that increases polynomially with the number of spin orbitals included in the active space.  Therefore exploring performance as a function of active space size is relevant for minimising the quantum resource required for extracting features. 
For that reason we repeat the analysis from Section~\ref{sec:motivational_results} with active space sizes (2e, 2o) and (6e, 6o) to see how the cross validation error from the resulting PLS model changes.

\begin{figure}[!hbt]
\includegraphics[scale=0.5]{"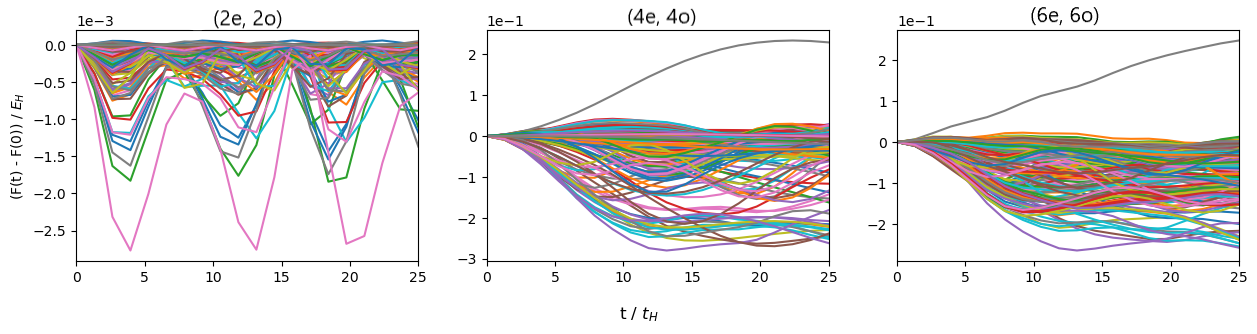"}
\caption{Features $F(t)$ forming the quantum fingerprint for different active space sizes from simulated unitary evolution from the fragment Hartree-Fock ground state with excitations. Different line colours correspond to different molecules.  Shown are the active spaces (2e, 2o) (\textit{left}), (4e, 4o) (\textit{middle}), and (6e, 6o) (\textit{right}).}
\label{fig:optimisations:active_space_features}
\end{figure}

\begin{figure}[!hbt]
\includegraphics[scale=0.6]{"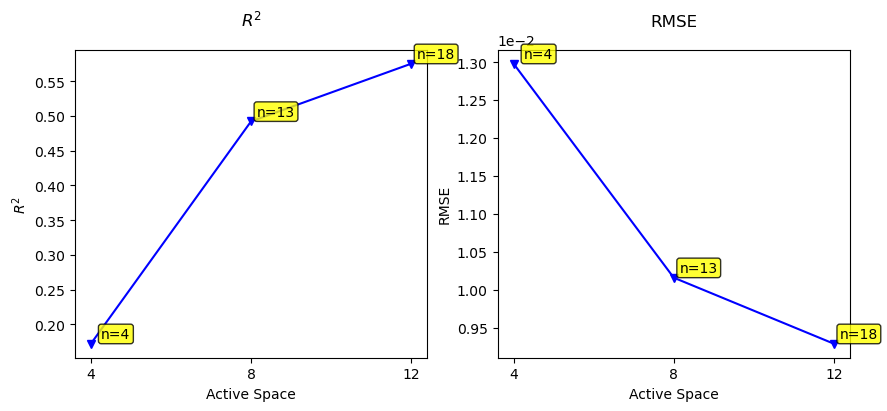"}
\caption{Cross-validated PLS performance metrics for different active space sizes for features captured from simulated unitary evolution from the fragment Hartree-Fock ground state with excitations.  Labelled are the number of components, $n$, selected in the PLS model.}
\label{fig:optimisations:active_space_pls}
\end{figure}

In Fig.~\ref{fig:optimisations:active_space_features}, the temporal trajectory of the observable $F(t)$ relative to its initial value $F(0)$ is shown.  This indicates that the (2e, 2o) active space exhibits dynamics which are completely inconsistent with larger active spaces (4e, 4o) and (6e, 6o).  This suggests a larger active space will likely to be required for useful predictive modelling.  This is confirmed in Fig.~\ref{fig:optimisations:active_space_pls} which shows the explained variance $R^2$ and root-mean-square-error (RMSE) when predicting the reactivity proxy based on molecular LUMO energies.  The performance of the predictive model is found to be poor for the (2e, 2o) active space. For the (6e, 6o) active space, the performance is found to increase beyond that found for the (4e, 4o) active space.

\FloatBarrier
\subsection{Initial state preparations}
\label{subsec:initial_states}
As discussed in the last section, the dynamics will be a function of the initial state on which the state transformation is applied. We therefore focussed our efforts on making it easier for our framework to incorporate different initial state configurations. In the Section~\ref{sec:pred_models}, the state transformation is unitary evolution from and excited state $\ket{\Phi^\text{ex}}$ defined in eq.~\ref{eq:initial_state}.
 
\begin{figure}[!hbt]
\includegraphics[scale=0.6]{"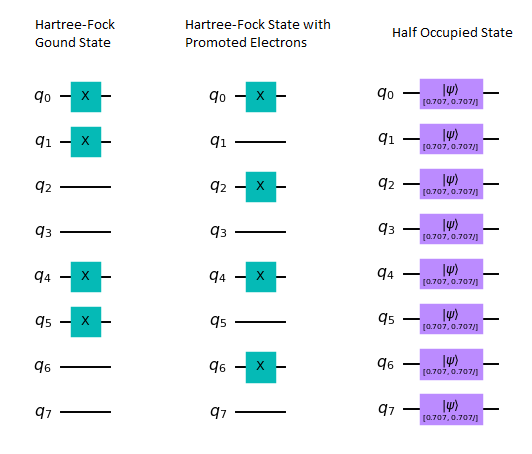"}
\caption{Circuit diagrams, representing three initial state implementations in our framework.  Shown \textit{left} is the 
Hartree-Fock ground state, \textit{centre} is the Hartree-Fock state with an excited election pair, and \textit{right} the half-occupied state created through a single-qubit rotation. In each case, this is for the (4e, 4o) active space.}
\label{fig:initial_state:circuit_diagrams}
\end{figure}

This initial state is constructed on a quantum computer by mapping each active spin orbital to a qubit, and applying an x-gate to the qubits corresponding to filled spin orbitals.  To explore different initial states, two additional initial state configurations were implemented.  Fig.~\ref{fig:initial_state:circuit_diagrams} summarises all three initial state circuit diagrams for an \sulf{} fragment described by (4e, 4o).

The first new initial state is the Hartree-Fock ground state for the fragment $\ket{\Phi_\text{HF}}$ which serves as another reference state for evolution.  The second new initial state we have explored is a `half-occupied' state.  This is prepared by applying a single-qubit rotation gate to each qubit leading to a superposition state where occupation probabilities are equal across spin orbitals in the (4e, 4o) active space. Note that the total number of electrons remains fixed across all three gate-efficient state preparation methods.  

\begin{figure}[!hbt]
\includegraphics[scale=0.4]{"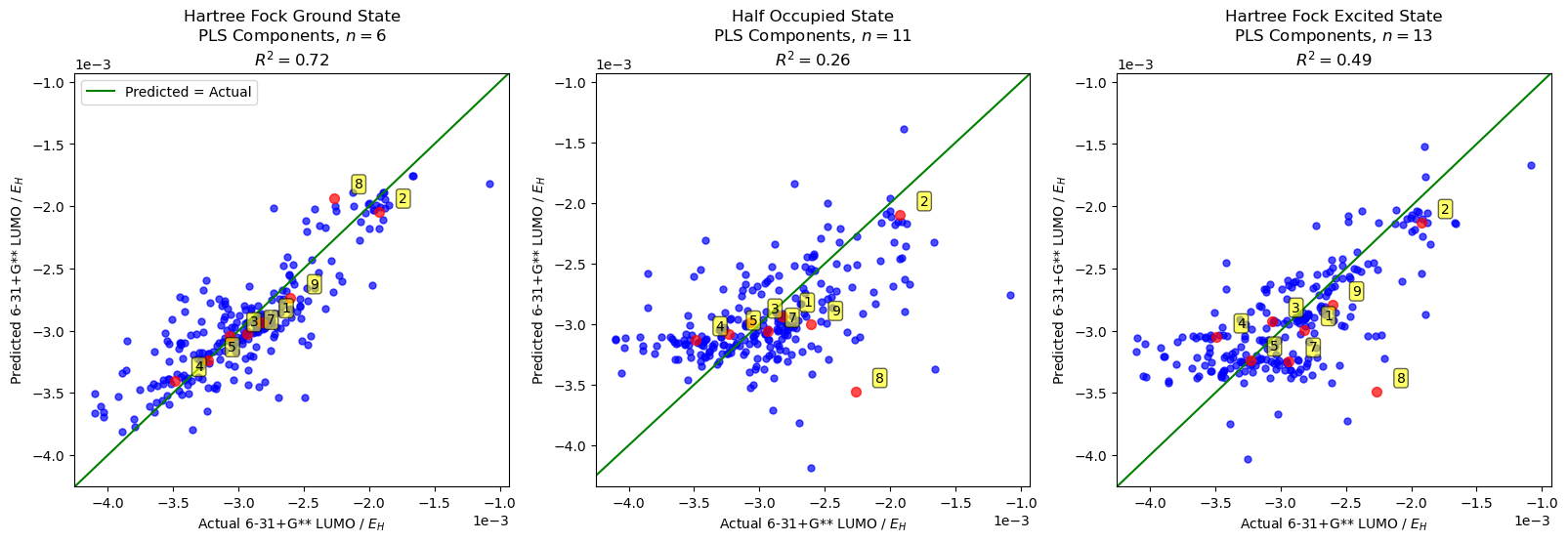"}
\caption{Results for each of the three initial states. Shown are the predictions for models trained each initial state described in Fig.~\ref{fig:initial_state:circuit_diagrams} for an active space (4e, 4o) and evolution time of $t/t_H = 25$.  Shown are cross-validation results for a 5-fold cross validation scheme.  The number of PLS components and explained variance $R^2$ is shown for each plot. Plot labels indicate the 8 molecules for which experimental reactivity data is available.}
\label{fig:initial_state:results}
\end{figure}

Results for the PLS model to predict the reactivity proxy are shown in Fig.~\ref{fig:initial_state:results}.  We find a significant improvement in predictive performance when moving from $\ket{\Phi^\text{ex}}$ to the Hartree-Fock ground state as the initial state.  The superposition state degrades performance below that of the $\ket{\Phi^\text{ex}}$ initial state.  We note that for different predictive modelling challenges with different target variables, the optimal initial state may be different.

\FloatBarrier
\subsection{Selection of measurement operators}
\label{subsec:measurements}
The results shown in Section~\ref{sec:motivational_results} focussed on measuring the projection of energy operators on to the \sulf{} fragment.  In this section, we extended the state measurement component of the workflow to facilitate an exploration of general one-body operators.  We note that while restricting to this class of operators in this work, the abstractions introduced in our framework allow for the support of more general cases such as two-electron operators, which will be useful for looking at electron-electron correlation functions. 

We write the general one-body operator as 

\begin{equation}
\expectation{O} = \sum_{rs} O_{rs} \expectation{\adag{}_{s} \ann{}_{r}}
\label{eq:observable}
\end{equation}

where $r,s$ index a general orthonormal basis of molecular orbitals, with $O_{rs} = \bra{\phi_{r}}\hat{O}\ket{\phi_{s}}$. Thus selecting an operator amounts to selecting a particular basis set of molecular orbitals $\ket{\phi_r}$ and the matrix coefficients of the operator in that basis. This flexibility allows us to explore two capabilities: firstly, it means that our measurements themselves can become data-driven in the sense that we can update our choice of measurement operator to give the quantum features that produce the best predictions for the particular down-stream machine learning task. This is discussed in the following section.

\FloatBarrier
\subsubsection{Data-driven measurements}
\label{subsec:pred_models:hyd}
To explore the idea of optimising the quantum state measurements performed we look at a toy example problem of predicting the inter-atomic distance between two Hydrogen atoms in a Hydrogen molecule as a contrived but illustrative example which does not require the use of calculated embeddings as we limit our study to the (2e, 2o) space. The inter-atomic distance is varied and the dynamics of single-electron observables used as a feature vector to train a machine learning model to predict atomic separation. A general one-body observable is calculated using eqs.~\ref{eq:evolution} and~\ref{eq:observable}. In this \hyd{} example, there are two fragment orbitals ($i=0,1$) yielding three matrix elements $O_{00}$, $O_{11}$ and $O_{01}=O_{10}$.
We prepare the \hyd{} molecule in the Hartree-Fock groundstate $\ket{\Phi_{\text{HF}}}$ evolve under Hamiltonian $H$ for different times, $t/t_H \in [0,4]$.  At each value of $t$ we measure $\expectation{O(t)}$. To explore this example, we created a dataset of 30 \hyd{} molecules with different inter-atomic distances $z$ ranging between $a_0$ and $3a_0$, where $a_0$ is the Bohr radius. Of these, 20\% were set aside randomly as a validation set and 10\% withheld as a final test set. The remaining 70\% were used for training a ridge regression model to predict $z$ from $\expectation{O(t)}$ at 8 values of $t$.

\begin{figure}[!hbt]
\includegraphics[scale=0.6]{"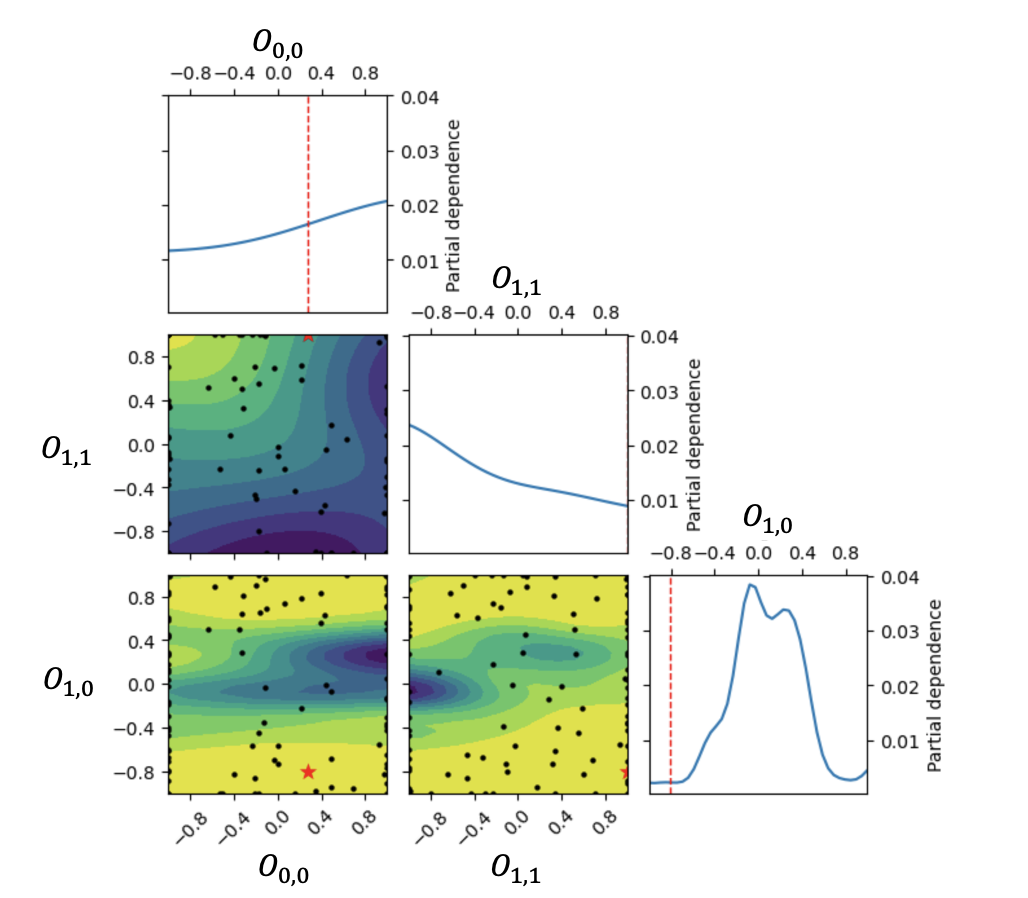"}
\caption{Features $\expectation{O(t)}$ for \hyd{} molecules with different interatomic distances $z$ from the test set, calculated using a noiseless quantum simulator for the optimal choice of quantum measurement.}
\label{fig:H2:opt}
\end{figure}

We used a kernel ridge regression model to predict the value of $z$ for a particular choice of the measured $\expectation{O(t)}$ allowing us to measure the validation mean square error.  This allows us to classically optimise the values $O_{00}$, $O_{11}$ and $O_{01}=O_{10}$ to minimise the validation error thus producing the optimal measurement for the downstream machine learning task.  While this example may seem trivial, it demonstrates how the framework allows the form of the quantum measurement can be driven by the results of a downstream machine learning problem. This not only helps to tune our method in a data-driven manner, but the form of the optimal measurement may uncover insights into the problem. Fig.~\ref{fig:H2:opt} shows the result of applying a Gaussian process to model the cross validation error
as a function of $O_{00}$, $O_{11}$ and $O_{01}=O_{10}$. The red star shows the optimal result $O_{00}\approx0.4$, $O_{11}\approx 0.8$ and $O_{01}\approx -0.8$, while the blue dots are points that are evaluated during the search to identify the optimal values.

\begin{figure}[!hbt]
\includegraphics[scale=0.3]{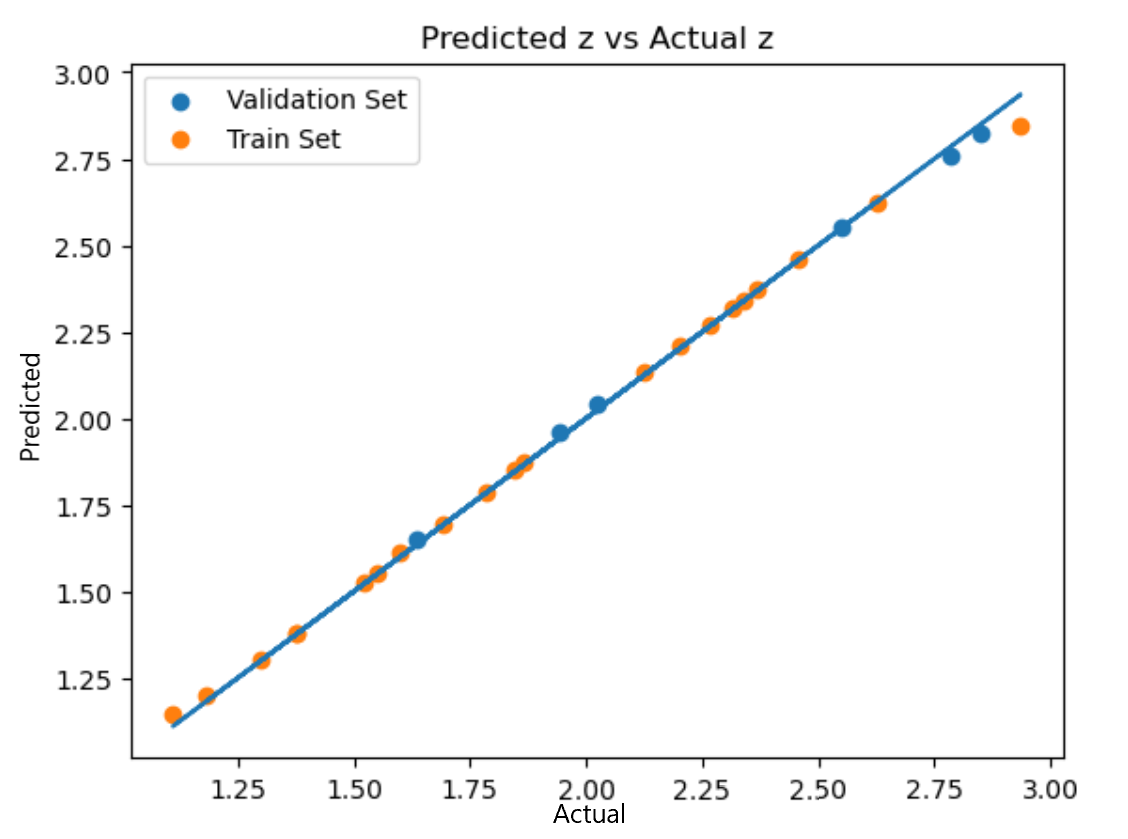}
\caption{Comparison of predicted $z$ vs actual $z$ distances for the dataset of \hyd{} molecules with different interatomic distances $z$. Shown are the training and validation data for the ridge regression model.}
\label{fig:H2:pred_vs_actual}
\end{figure}

\begin{figure}[!hbt]
\includegraphics[scale=0.6]{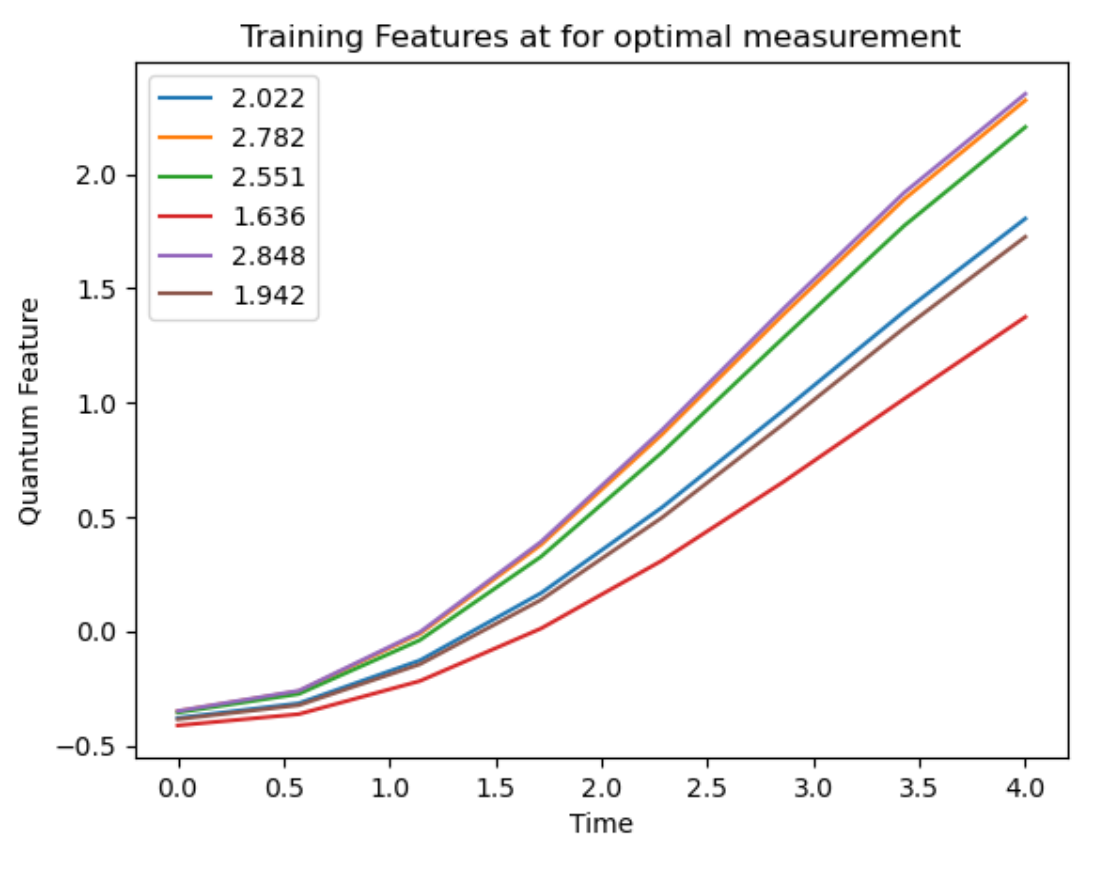}
\caption{Features $\expectation{O(t)}$ for \hyd{} molecules with different interatomic distances $z$ from the test set, calculated using a noiseless quantum simulator.}
\label{fig:H2:features}
\end{figure}

In Fig.~\ref{fig:H2:pred_vs_actual} we plot the mean square error for both the training and validation set showing good agreement and in Fig.~\ref{fig:H2:features} we plot the quantum features for the different values of $z$ in the validation set. These highlight that the dynamics allow for separation of molecules with different $z$. The results of the ridge regression model used to predict $z$ are shown in Fig.~\ref{fig:H2:pred_vs_actual}, demonstrating that the approach allows an accurate predictive model for inter-atomic separation $z$.

\FloatBarrier

\FloatBarrier
\section{Quantum algorithms}
\label{sec:quantum_algos}

In the previous sections we have explored some questions that relate to running the pipeline on real quantum hardware, such as duration of evolution and active space size.  In this section we look at running the pipeline on quantum hardware.  To run Hamiltonian simulation on quantum hardware we must use a quantum circuit that approximates $e^{-iHt}$. We study different methods to approximate this unitary on NISQ hardware, which leads us to a discussion of more general unitary transformations $U_t(H)$ which could be used to extract features for a quantum fingerprint. This section primarily focusses on the Trotter-Suzuki approach before exploring the feasibility of the variational fast-forwarding method~\cite{Cirstoiu2020} in the final subsection.

The Trotter-Suzuki approach approximates the unitary evolution operator as a product of non-commuting terms. Note that eq.~\ref{eq:H_DMET} can be expressed in a general form $H = \sum_{j=1}^m H_j$ where the $m$ terms $H_j$ do not commute with each other.  It is then possible to decompose the time evolution operator as
\begin{equation}
e^{-iHt} = \left(\prod_{j=1}^m e^{-\frac{iH_j t}{r}}\right)^r + \mathcal{O}\left(\frac{m^2 t^2}{r}\right)
\end{equation}
which can be systematically improved by constructing exponentials which cancel the error terms.  To second order, the Trotter-Suzuki formula is
\begin{equation}
e^{-iHt} = \left(\prod_{j=1}^m e^{-\frac{iH_j t}{2r}} \prod_{j=m}^1 e^{-\frac{iH_j t}{2r}} \right)^r + \mathcal{O}\left(\frac{m^3 t^3}{r^2}\right)
\label{eq:2nd_order_trotter}
\end{equation}

We can select the order of the product formula and the number of repetitions $r$ to optimise for the error of the approximation, noting that for given $H$ and $t$ on NISQ experiments, there will be a trade-off with bigger $r$ resulting increasing the number of gates.

\FloatBarrier
\subsection{Shallow approximations to Hamiltonian simulation}
\begin{figure}[!hbt]
\includegraphics[scale=0.8]{"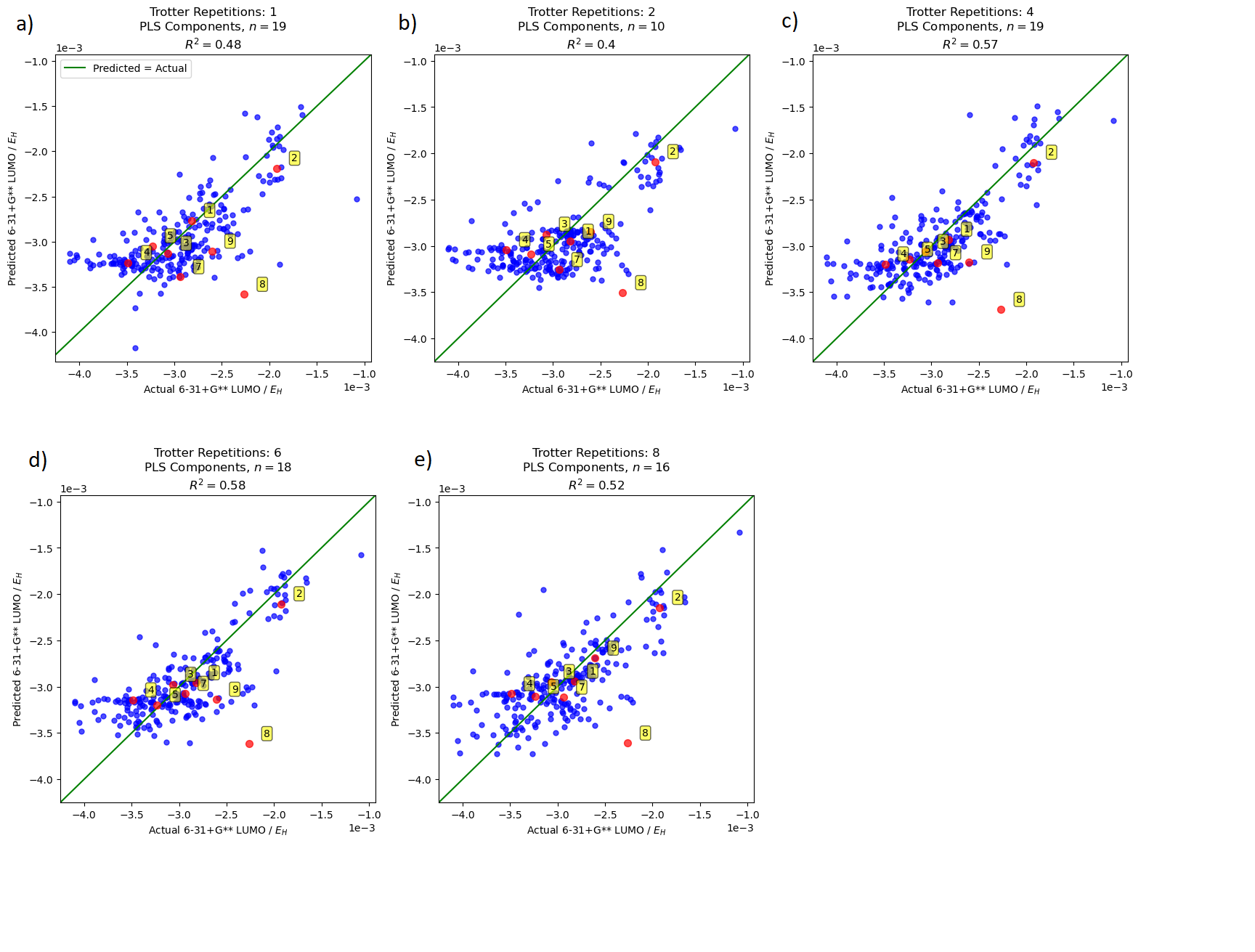"}
\caption{Cross-validated results for PLS models for quantum features extracted from evolution under the Trotter-Suzuki method with different number of repetitions $r$, with: a) $r=1$, b) $r=2$, c) $r=4$, d) $r=6$ and e) $r=8$.  Shown are the predictions for models trained each initial state described in Fig.~\ref{fig:initial_state:circuit_diagrams} for an active space (4e, 4o) and evolution time of $t/t_H = 25$.  Shown are cross-validation results for a 5-fold cross validation scheme.  The number of PLS components and explained variance $R^2$ is shown for each plot. Plot labels indicate the 8 molecules for which experimental reactivity data is available.  }
\label{fig:trotter_reps}
\end{figure}

There is no reason {\it a priori} why the transformation applied to initial states for feature extraction needs to be the exact simulation of the temporal evolution under $H$.  Therefore we can consider a range of state transformations $U_t(H)$ based on different approaches and levels of approximation to Hamiltonian simulation. In Fig.~\ref{fig:trotter_reps} we demonstrate the predictive performance of our PLS model when we approximate the Hamiltonian simulation using different numbers of Trotter repetitions.

Interestingly, we note that modest performance is found even for the case of a single Trotter repetition $r=1$, despite the general -- albeit non-monotonic -- trend towards better performance for increasing $r$.

\FloatBarrier
\subsection{Quantum Hardware}

The Trotter-Suzuki approach is explored in the context of the \hyd{} toy example discussed in Section~\ref{subsec:pred_models:hyd}.  We focus on two distinct examples, Molecules 1 and 2, with respective inter-atomic distances $1.2a_0$ and $2a_0$, and compare the results of generating the temporal feature vector -- the quantum fingerprint -- from a noiseless quantum simulator and a quantum circuit running on quantum hardware via the IBM Qiskit Runtime with different error mitigation techniques applied. 

\begin{figure}[!hbt]
\includegraphics[scale=0.7]{"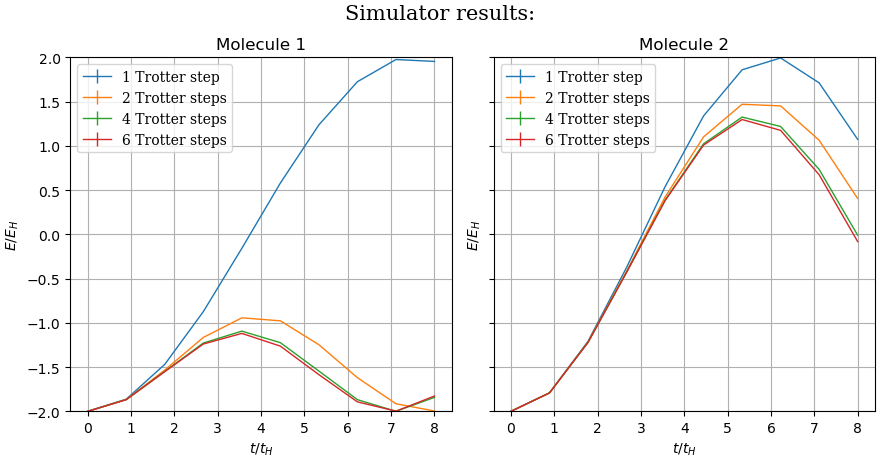"}
\caption{Results from the Trotter-Suzuki expansion with different numbers of repetitions for \hyd{} molecules 1 and 2, respectively with interatomic distances 1.2$a_0$ and 2$a_0$. }
\label{fig:quantum_hardware:trotter_test}
\end{figure}

We first explore the second-order approach in eq.~\ref{eq:2nd_order_trotter} with different numbers of repetitions $r$, for the two bond lengths considered, on a noiseless quantum simulator.  The results are shown in Fig.~\ref{fig:quantum_hardware:trotter_test}, where it is clear that a single repetition $r$ is insufficient to reproduce distinct temporal trajectories of $\expectation{O(t)}$, but rapid convergence is seen across all times up to $t=8$ for $r \ge 2$.

\begin{figure}[!hbt]
\includegraphics[scale=0.7]{"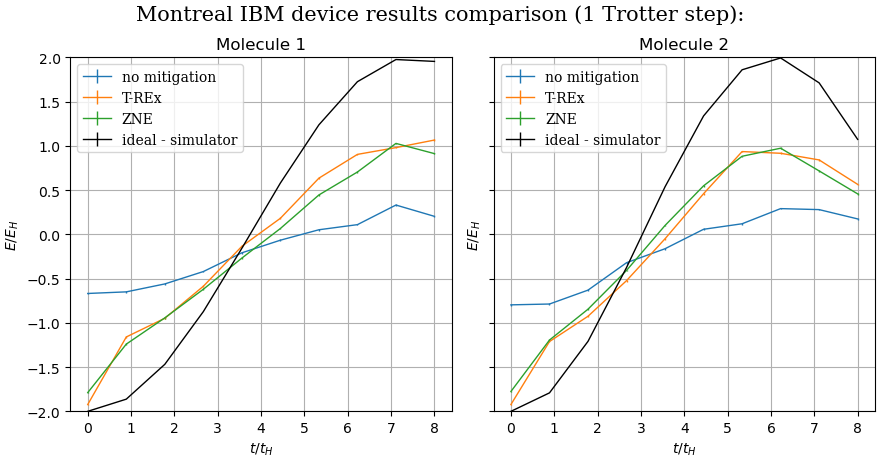"}
\caption{Quantum fingerprint results from the Trotter-Suzuki expansion with $r=1$ for \hyd{} molecules 1 and 2, respectively with interatomic distances 1.2$a_0$ and 2$a_0$, for noiseless statevector simulation, quantum hardware with no error mitigation, and the ZNE and T-REx error mitigation approaches.}
\label{fig:quantum_hardware:trotter_compare_r1}
\end{figure}

\begin{figure}[!hbt]
\includegraphics[scale=0.7]{"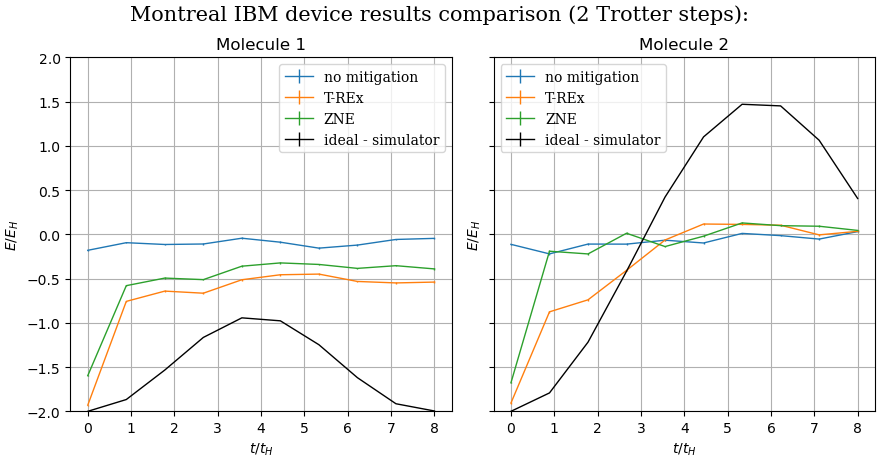"}
\caption{Quantum fingerprint results from the Trotter-Suzuki expansion with $r=2$ for \hyd{} molecules 1 and 2, respectively with interatomic distances 1.2$a_0$ and 2$a_0$, for noiseless statevector simulation, quantum hardware with no error mitigation, and the ZNE and T-REx error mitigation approaches.}
\label{fig:quantum_hardware:trotter_compare_r2}
\end{figure}

Crucially, Fig.~\ref{fig:quantum_hardware:trotter_test} shows that it is possible to distinguish the two molecules from their quantum fingerprints when $r=2$, and may be possible for $r=1$ despite the poor approximation to $e^{-iHt}$.  We therefore take the Trotter-Suzuki formula with $r=1$ and $r=2$ and explore the quantum simulator results as well as the IBM Montreal quantum device~\cite{IBM}. Figs.~\ref{fig:quantum_hardware:trotter_compare_r1} and~\ref{fig:quantum_hardware:trotter_compare_r2} show the $r=1$ and $r=2$ results, respectively, for different error mitigation approaches as well as the for the case with no error mitigation strategy.  Results from quantum hardware are degraded in comparison to the quantum simulator without noise.  

Due to the longer circuit depth, the results for $r=2$ show less difference in the quantum fingerprints between the two molecules, indicating that a data-driven approach would be more effective for a single repetition.  This suggests that a more general $U_t(H)$ optimised to minimise the quantum resource required could be more effective for predictive modelling, despite being a significantly worse approximation to $e^{-iHt}$. 

\begin{figure}[!hbt]
\includegraphics[scale=0.55]{"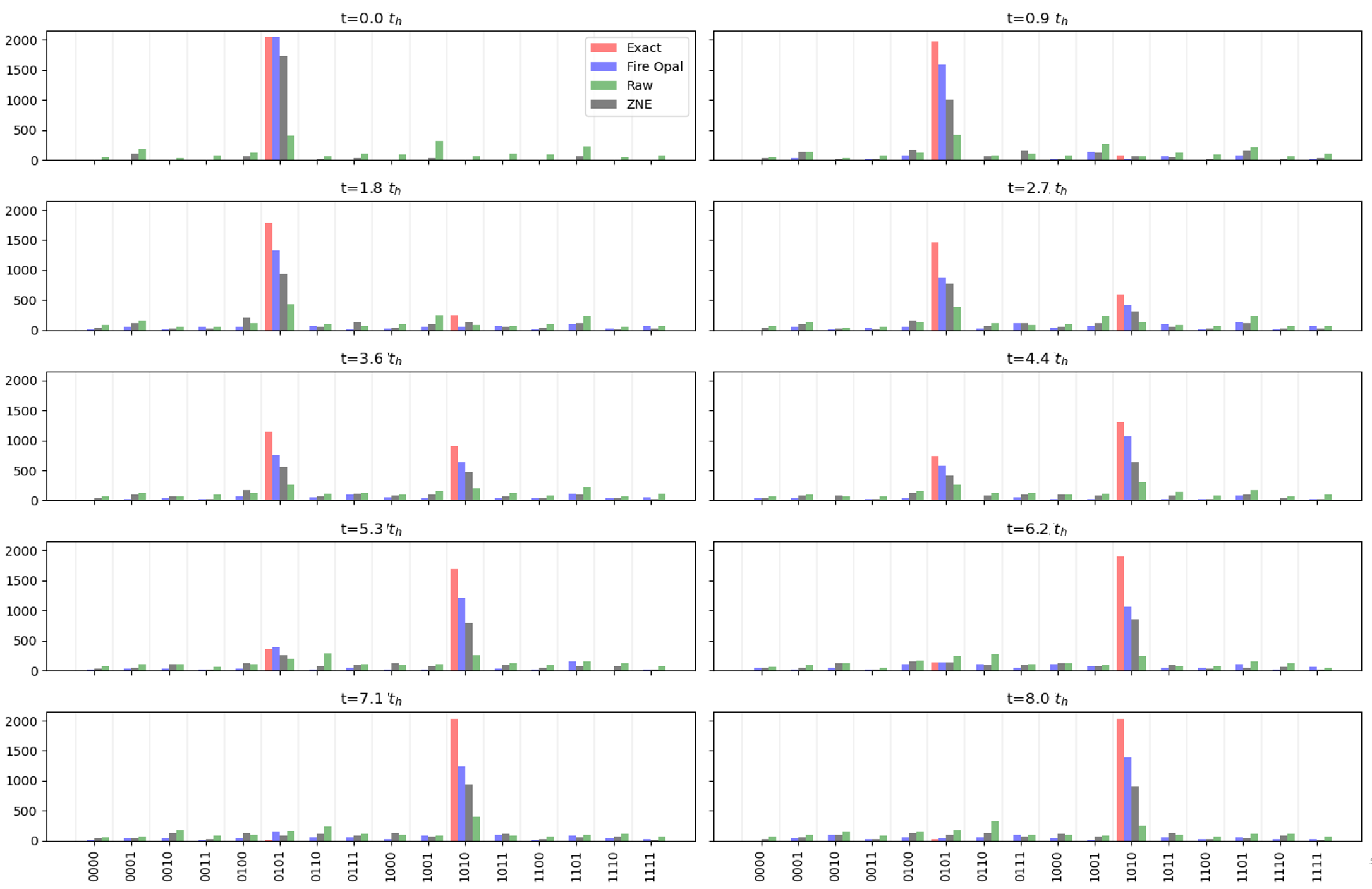"}
\caption{Measurement histograms for evolution with a single Trotter step, comparing results from exact evolution on a quantum simulator, quantum hardware without error mitigation ('raw'), resulting counts using Fire Opal, and results using IBM's implementation of zero noise extrapolation (ZNE) for comparison.}
\label{fig:quantum_hardware:fire_opal}
\end{figure}

Finally, we present results obtained using the Fire Opal package for automated error suppression~\cite{Mundada2022}.  Histograms of measurements from different evolution times are shown in Fig.~\ref{fig:quantum_hardware:fire_opal}.  Exact results from a quantum simulator and results from quantum hardware with and without error mitigation are shown for comparison.  This indicates that Fire Opal performs even better than zero noise extrapolation.  We anticipate that a combination of approaches will yield even better results due to the complementary nature of the methods.

\FloatBarrier
\section{Conclusions}
In this work, we have presented an end-to-end predictive modelling pipeline using features forming a {\it quantum fingerprint} which will be calculable on future quantum hardware.  The pipeline is flexible, allowing for calculation of quantum features on different compute back ends with different strategies for error suppression and mitigation.  It is further possible to customise to a range of challenges.  In particular, we have focussed on exploring the three principal steps in the pipeline associated with generating the quantum fingerprint -  initial state preparation, algorithms for state transformation by Hamiltonian simulation, and measurement strategies. 

We explored the method on the challenge of predicting the reactivity of molecules in a series with sulfonyl fluoride covalent warheads through using a DMET embedding for the \sulf{} group.  We established that the pipeline is able to predict a proxy for molecular reactivity calculated from DFT simulations for the entire molecule.  Our method used a significantly smaller active space than the prior DFT simulations, and used features from the \sulf{} fragment captured from temporal evolution of the many-body fragment Hamiltonian.  The features were used to train a partial least squares model, with performance assessed by cross validation.  We focussed on a range of scenarios for feature extraction.  In particular we found the performance increased for larger active spaces and longer evolution times. Lower-order approximations to Hamiltonian simulation (requiring shorter circuit depths) were however found still to be predictive.  This analysis suggests the potential for utility on future hardware.  

We now discuss what our findings suggest as the best next  steps.  In the short term, there are some practical steps we can take that build on the pipeline to make it more applicable to current and near-term NISQ devices. The first approach is to note that it may be possible to look at different classes of unitaries for state transformation, $U_t(H)$, that can be more efficiently encoded on quantum computers~\cite{Kim2023}; that is to say that the choice of $U_t(H) = e^{iHt}$ need not be the only type of state transformation which captures similarities and differences between structures at the quantum many-body level. Secondly, we note that the choice of measurement operators to create a quantum fingerprint could be parametrised, with the parameters learnt through optimisation on training data. This further opens the door to quantum machine learning techniques being used within a data-driven pipeline similar to the one studied in this work. 

\section*{Acknowledgements}
We are grateful to Franziska Wolff, Phalgun Lolur and Julian van Velzen for insightful discussions.  Part of this work was funded under an STFC Cross-Cluster Proof of Concept Grant and Highlight Call on Quantum Computing in collaboration with the NQCC (Ref.~POC2022-Q2).

\bibliographystyle{apsrev4-2}
\bibliography{report}
\end{document}